\documentclass[twocolumn, showpacs, aps] {revtex4}
\ifx\pdfoutput\undefined
  \usepackage[dvips]{graphicx}
\else
  \usepackage[pdftex]{graphicx}
\fi
\usepackage{amsmath}
\usepackage{latexsym}

\begin{document}
\draft
\title{Radio-frequency-driven motion of single Cooper pairs
across the superconducting single-electron transistor with
dissipative environment}

\author{A.~B.~Zorin}

\author{S.~V.~Lotkhov}
\affiliation{Physikalisch-Technische Bundesanstalt, 38116
Braunschweig, Germany}

\author{S.~A.~Bogoslovsky}
\altaffiliation[Permanent address:~] {Laboratory of
Cryoelectronics, Moscow State University, 119899 Moscow, Russia}

\author{J.~Niemeyer}
\affiliation{Physikalisch-Technische Bundesanstalt, 38116
Braunschweig, Germany}


\begin{abstract}

We report on the effect of the frequency-locked transfer of single
Cooper pairs in a superconducting single-electron Al transistor
embedded in a dissipative environment (on-chip Cr resistor of $R
\approx 40$ k$\Omega$). The transistor was dc voltage biased, and
the harmonic signal of frequency $f$ of several MHz was applied
to the gate. Due to the substantial rate of relaxation, the
unidirectional transfer of single pairs occurred in each junction
once per clock cycle and the current plateaus at $I\approx 2ef$
were developed in the transistor's $I-V$ curves. The mechanisms
(supercurrent, Landau-Zener tunneling, quasiparticle tunneling,
etc.) deteriorating the phase-locking regime are discussed.
\end{abstract} 

\maketitle

\section{Introduction}

In the last few years, the attention of experimenters has been
attracted to superconducting systems of small tunnel junctions
whose behavior benefited from an interplay of Coulomb and
Josephson effects \cite{Geerligs, Matters, Haviland, Kuzm, Flees,
Nakamura, BBands, Rimberg, Bouchiat, QBit, Penttila, Bl-electr,
CP-Pump}. Besides the interesting physical aspects, such as the
charge-phase duality \cite{Matters}, Cooper-pair solitons
\cite{Haviland}, Bloch-band structure
\cite{Flees,Nakamura,BBands}, the effect of dissipation
\cite{Kuzm, Rimberg, Penttila}, etc., the systems of small
Josephson junctions may also be useful for applications like
Cooper-pair electrometers \cite{Bl-electr}, pumps \cite{Geerligs,
CP-Pump}, charge qubit \cite{Bouchiat,QBit}, etc. In the light of
fundamental metrology and quantum computation, the researchers
are particularly interested in the controlled transfer of
individual pairs across a superconducting circuit
\cite{MKeller,AverinQC}.

In this paper we focus on the single pair transport in the
two-junction circuit with a gate, or, as it is sometimes called,
the Bloch transistor \cite{AvLikh}. The behavior of this system
is determined by the relationship between the Coulomb energy
$E_{\rm{c}}$ and the Josephson coupling energy $E_{\rm{J}}$. In
particular, if the Coulomb energy dominates, $E_{\rm{c}} \gg
E_{\rm{J}}$, the transfer of pairs is a resonance process. The
motion of pairs across the transistor can be realized in three
different ways: (i) by means of the through-supercurrent at the
bias voltage $V \approx 0$ and the gate-induced charge on the
island $Q \approx \pm e$, (ii) by Josephson tunneling or/and an
Andreev reflection process in one junction accomplished by
quasiparticle tunneling in another junction at $V\neq0$ and at
certain values of $Q$ (see Refs. \cite{Fulton,Nakamura,Alec} and
\cite{Fitz}, respectively), and (iii) by means of autonomous
periodic tunneling of single pairs (Bloch oscillations)
\cite{LiZo,HavPashKuz}, this regime being possible in a very
high-ohmic electromagnetic environment ($R \gg R_{\rm{Q}}\equiv
h/4e^2 \approx 6.5$ k$\Omega$, the resistance quantum) or, in
other terms, at the dc current bias.

As long as the phase across the transistor in the voltage-bias
regime is meaningful, the process (i) yields the current with the
\textit{meaningless} number of pairs tunneling per unit time
across the device. Either of the mechanisms (ii) gives the
\textit{random} number of the particles transferred. Although the
current-bias regime (iii) should ensure the \textit{coherent}
motion of pairs, in the experiment, the resulting current is
strongly affected by fluctuations \cite{HavPashKuz}.

The goal of this paper is to realize a regime of the gate-driven
unidirectional motion of single Cooper pairs in the voltage bias
transistor. Since Cooper pair tunneling is by its nature an
elastic, and, hence, reversible process, the desired regime, as
we will show below, can be achieved by introduction of finite
damping. Endorsing this idea we fabricated an Al transistor with
on-chip resistor of $R\sim $ several $R_{\rm{Q}}$, connected in
series to the transistor. In the $I-V$ curves of this device we
observed remarkable current plateaus at $I=2ef$, reflecting the
transfer of one Cooper pair per cycle of harmonic signal of
frequency $f$ applied to the gate.

\section{Circuit parameters}

\begin{figure}
\begin{center}
\leavevmode
\includegraphics[width=2.8in]{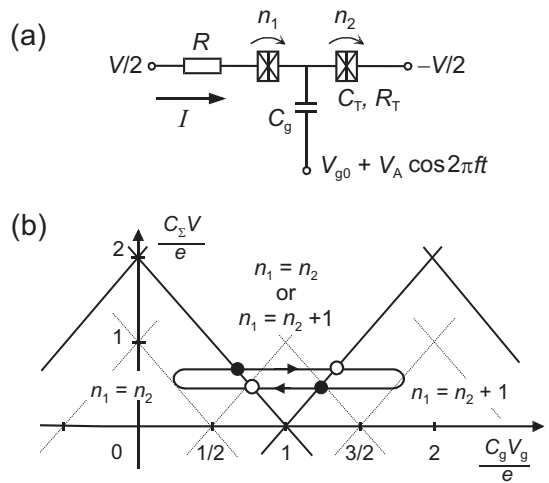}
\caption {(a) Equivalent circuit of the device. The crossed-box
symbols denote the Josephson junctions with small capacitance.
The device is driven by a harmonic signal via the gate
capacitance. (b) The boundaries of the states with fixed number
of Cooper pairs on the island (shown by solid lines) and the
boundaries for single-electron tunneling (shown for comparison by
dashed lines.) For the sake of clarity, the cycle trajectory
(note, that $V=$~const) is shown as the squeezed loop. The solid
(open) circles mark the processes accompanied by the single-pair
(zero) charge transferred across the corresponding junction.}
\label{EqvSchm}
\end{center}
\end{figure}

The diagram of the electric circuit comprising two small Josephson
junctions with a capacitive gate and a resistor attached is
depicted in Fig.~\ref{EqvSchm}a. The system parameters are
assumed to be:
\begin{equation}
\label{Energies} E_{\rm{c}} \gg E_{\rm{J}} \gg k_{\rm{B}}T,
\end{equation}
where $E_{\rm{c}}=e^2/2C_\Sigma$ with
$C_\Sigma=C_1+C_2+C_{\rm{g}}$, denoting the total capacitance of
the transistor island including the capacitances of individual
junctions $C_{1,2}=C_{\rm{T}}$ and the gate capacitance
$C_{\rm{g}} \ll C_{\rm{T}}$. The Josephson coupling energy in
either junction is $E_{\rm{J}}=(\Phi_0/2\pi) I_{\rm{c0}} =
(R_{\rm{Q}}/2R_{\rm{T}}) \Delta$, where $\Phi_0=h/2e \approx
2.07$~fWb is the flux quantum, $I_{\rm{c0}}$ and $R_{\rm{T}}$ are
the critical current and the tunnel resistance of the individual
junctions, respectively, $\Delta$ is the superconductor energy gap
and $T$ the temperature.

The effect of the external resistor (which is characterized by the
dimensionless parameter $z = R/R_{\rm Q}$) can be described as
follows: First, the resistor "hampers" the tunneling of pairs in
a particular junction. As a result, tunneling occurs when the
change of the energy associated with this tunneling is positive,
$E>0$; the energy $E$ dissipates in the resistor. In the limit of
$\lambda \equiv E_{\rm{J}}/E_{\rm{c}} \rightarrow 0$, the rate of
the one-pair tunneling \cite{AvOd,IngNaz} is
\begin{equation}
\label{Gamma} \Gamma(E)=\frac{\pi}{2\hbar} E_{\rm{J}}^2 P(E),
\end{equation}
where $P(E)$ is a peak-shaped "environment" function obeying the
normalization condition $\int_{-\infty}^{+\infty}P(E){\rm d}E =1$
and the relation $P(E) \propto E^{2z'-1}$, at small positive $E$.
The parameter $z'=\frac{1}{4}z$, where the factor of
$\frac{1}{4}=(C_{\rm{T}}/C_\Sigma)^2$ accounts for the reduction
of the damping effect due to the capacitance of the neighboring
junction connected in series to the resistor $R$ (see circuit
analysis in Refs.\cite{IngNaz,IngWyr}).

Secondly, the resistor suppresses through-tunneling across the
whole transistor or, in other terms, the total supercurrent
$I_{\rm{s}}$. (Note, at $R = 0$ and $\lambda \rightarrow 0$, the
total critical current $I_{\rm{c}} = I_{\rm{c0}}/2$
\cite{AvLikh}.) That is to say that for the most intensive
(resonance) tunneling occurring at $Q \approx \pm e$, the
supercurrent is expressed as
\begin{equation}
\label{GammaTot} \frac{I_{\rm{s}}(V)}{I_{\rm{c}}}=\frac{\pi}{8}
E_{\rm{J}}P_{\rm{tr}}(2eV) \approx 0.02 \lambda
\left(\frac{4V}{V_0}\right)^{2z-1},
\end{equation}
where $P_{\rm{tr}}(E)$ is the environment function for the whole
transistor and $V_0 = e/C_{\Sigma}$. The second relation in
Eq.(\ref{GammaTot}) is valid for small $V$, viz., $0<V \ll V_0/4$
and $z>1/2$. In order to ensure sufficient suppression of the
supercurrent and, at the same time, to have sufficiently large
$\Gamma(E)$ at small $E$, the compromise values for $R$ of about
several $R_{\rm{Q}}$ should be chosen.

Finally, we assume that the rate of single electron
(quasiparticle) tunneling across the transistor junctions is
negligibly small. The factors which lead to its rate being
suppressed are: a high pregap resistance $R_{\rm{qp}}(V) \gg
R_{\rm{T}}$ at voltages $|V| \leq 2\Delta/e$, the presence of a
finite resistance $R$ in the external circuit \cite{IngNaz}, and
(at $E_{\rm{c}}<\Delta$) the even-odd parity effect
\cite{Tuominen}.

\section{Operating principle}

The circuit is dc voltage biased and the harmonic signal
$V_{\rm{g}}(t)$ of frequency $f$, amplitude $V_{\rm{A}}$ and dc
offset $V_{\rm{g0}}$ is applied to the gate, which leads to the ac
polarization of the island. Figure~\ref{EqvSchm}b shows the
working cycle in the $\{V_{\rm{g}},V\}$ plane. The cycle
trajectory hits the boundaries corresponding to the resonant
tunneling of pairs (at $R=0$) through the first and second
junctions. At these boundaries the values of the electric energy
with and without one extra pair on the island are equal, i.e.,
\begin{equation}
\label{Resonances}E_{n_1,n_2}=E_{n_1\pm 1,n_2}\quad {\rm
and}\quad E_{n_1,n_2}=E_{n_1,n_2\pm 1}.
\end{equation}
Here $n_1$ ($n_2$) denotes the number of pairs having traversed
the first (second) junction; the energy, including the work done
by the voltage source, reads \cite{AvLikh}
\begin{equation}
\label{En12}E_{n_1,n_2}=
4E_{\rm{c}}[(v_{\rm{g}}+n_1-n_2)^2-(n_1+n_2)v],
\end{equation}
where $v_{\rm{g}}=C_{\rm{g}}V_{\rm{g}}/2e = Q/2e$ and $v=V/2V_0$.

\begin{figure}
\begin{center}
\includegraphics[width=2.8in]{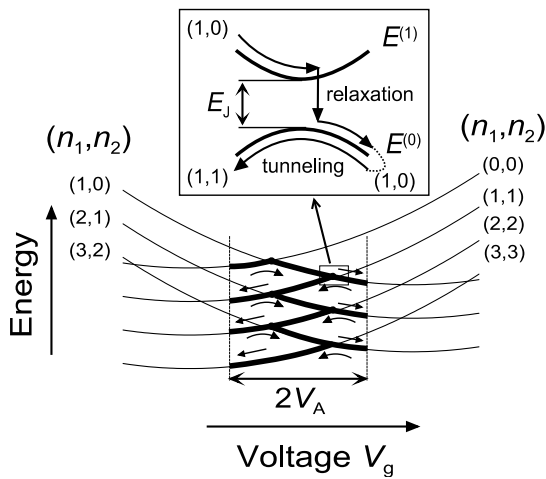}
\caption{The cycle trajectory (thick solid line) on the
$\{V_{\rm{g}},E_{n_1,n_2}\}$ plane. The arrows indicate the
direction of motion. The inset shows in detail the avoided
crossing region; two processes are schematically shown: (1)
relaxation $E^{(1)} \rightarrow E^{(0)}$, occurring without
transfer of charge (open circles in Fig.1b) and (2) motion along
the lower branch $E^{(0)}$, accompanied by single Cooper pair
tunneling (solid circles in Fig.1b).} \label{OpPrinc}
\end{center}
\end{figure}

Owing to the finite Josephson coupling, the charge states in the
vicinity of resonances, Eq.(\ref{Resonances}), are
quantum-mechanically mixed. As a result, e.g., for the states
$(n_1,n_2)$ and $(n_1+1,n_2)$, the total energy of the system
assumes the values \cite{LiZo}:
\begin{equation}
\label{Esplit}E^{(0,1)}={\rm const}\mp
\sqrt{(E_{n_1,n_2}-E_{n_1+1,n_2})^2+E_{\rm{J}}^2/4},
\end{equation}
where the lower $E^{(0)}$ and upper $E^{(1)}$ Bloch-band states
are separated by the energy gap of width $=E_{\rm{J}}$. Thus, our
system is mapped onto the two-level crossing problem.

In the absence of dissipation ($R=0$), a slow passage of the
avoided crossing region along either branch, Eq.(\ref{Esplit}), in
either direction leads to one pair passing in a certain direction;
these processes are reversible. At a larger rate of motion, the
interband transitions $E^{(0)} \rightarrow E^{(1)}$ and $E^{(1)}
\rightarrow E^{(0)}$ may occur; they are not accompanied by pair
transfer. The probability of such an event is given by the
Landau-Zener formula \cite{LandZen},
\begin{equation}
\label{LandauZener}p_{\rm LZ} = \exp(-\gamma)\quad {\rm
with}\quad \gamma={\frac{\pi E_{\rm{J}}^2}{2\hbar \dot{E}}}\,,
\end{equation}
where the derivative $\dot{E} = |\frac{{\rm} d}{{\rm
d}t}(E_{n_1,n_2}-E_{n_1+1,n_2})| \propto f$. If the cycling
frequency $f$ is low enough, $\exp(-\gamma) \ll 1$ and the
probability of the Cooper pair transfer $p_{\rm CP} = 1 - p_{\rm
LZ} \approx 1$.

In the case of finite dissipation and at sufficiently low
temperature $T$, relaxation from state $E^{(1)}$ to $E^{(0)}$ may
occur during the passage of the avoided crossing region starting
in the upper branch. For the appreciable strength of dissipation,
according to the theory of Ao and Rammer \cite{Ao} describing the
dynamics of the two-state system in a dissipative environment,
the probability of relaxation is
\begin{equation}
\label{Relax} p_{\rm rel} = 1-\exp(-\gamma)+\exp(-2\gamma) \approx
1
\end{equation}
with $\gamma$ given by the second expression in
Eq.(\ref{LandauZener}). In our notations this result is valid for
$z'>1$ (or $R>4R_{\rm Q}$) and $k_{\rm B}T < \hbar/RC_\Sigma$.
Note that with the motion within the lower branch the probability
of the transition $E^{(0)} \rightarrow E^{(1)}$ due to the
Landau-Zener mechanism remains unchanged (see
Eq.(\ref{LandauZener})), i.e. the probability $p_{\rm CP}\approx
1$ \cite{Ao}.

Figure~\ref{OpPrinc} shows the cycle trajectory in the energy
representation: the system glides along the pieces of parabolas,
Eq.(\ref{En12}). The inset is a blow-up of the avoided crossing
region in which first the relaxation (giving the zero charge
transferred) and then the crossing of the energy maximum (giving
the transfer of charge of $2e$) occur. As a result of the
back-and-forth sweep, one pair traverses both junctions once per
cycle, carrying the average current $I=2ef$. The energy
dissipation occurring in the electric circuit (in resistor $R$)
is equal to $2eV$ per cycle.

\section{Experiment}

The tunnel structures of type Al/AlO$_x$/Al (of nominal
dimensions 40~nm by 40~nm) with Cr microstrip resistor (with sizes
0.1~$\mu$m$\times 7~\mu$m$ \times 7~$nm) were fabricated by the
three angle evaporation technique described elsewhere \cite{Towa,
Pump}. The tunnel resistance of the junctions $R_{\rm{T}}$ was
around 35~k$\Omega$ and, assuming the experimental value of
$\Delta_{\rm Al}=175\mu$eV, this yielded $E_{\rm{J}} \approx
16~\mu$eV.

From the measurements of the $I-V$ and $I-V_{\rm g}$
characteristics in the normal state (namely, in magnetic field of
induction $B=1$~T), the capacitances of the junctions
$C_{\rm{T}}$ and of the gate $C_{\rm{g}}$ were found to be about
160~aF and 12~aF, respectively. The total capacitance of the
island $C_{\Sigma}$ and the charging energy $E_{\rm{c}}$ were
about 350~aF and 270~$\mu$eV, respectively. The ratio of the
characteristic energies in the sample thus was equal to
$\lambda\approx 0.06$. The resistance of the Cr strip was
evaluated at $R=40~k\Omega$, so the dimensionless parameter
$z\approx 6$.

\begin{figure}
\begin{center}
\leavevmode
\includegraphics[width=3.0in]{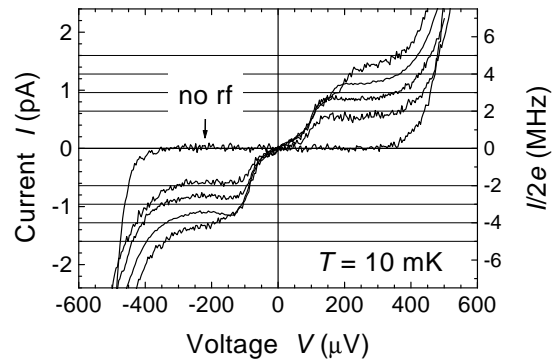}
\caption {The $I-V$ curves of the sample without rf drive and
driven by harmonic signals of frequency $f=2, 3, 4,$ and 5~MHz
with amplitude $V_{\rm{A}} \approx e/C_{\rm{g}}\approx 13$~mV.
The dc component of the gate voltage $V_{\rm{g0}}$ was tuned to
maximize the sizes of the plateaus. The right axis converts the
current values into the frequency units.} \label{IVcurves}
\end{center}
\end{figure}

The dc and rf measurements were carried out in a dilution fridge
at the bath temperature $T = 10$~mK. The recorded $I-V$ curves
are shown in Fig.~\ref{IVcurves}. The Coulomb-blockade type of
the autonomous characteristics showed, in particular, that the
critical current of the transistor was dramatically suppressed.
Under the action of the rf signal applied to the gate, the $I-V$
curve changed its shape: the blockade was lifted and the
characteristic plateaus were formed close to the current levels
$I=\pm 2ef$. The largest plateaus were recorded at the lowest
frequencies of the drive, i.e., at $f=2$ and 3~MHz.

\section{Discussion}

The electric current in this transistor is obviously associated
with the gate-driven motion of individual Cooper pairs.
Furthermore, this preliminary experiment even demonstrates some
superiority of the device characteristics over those of the
three-junction Cooper pair pumps \cite{Geerligs, CP-Pump}. On the
other hand, there are several factors which affect the shape and
the height of the plateaus. These factors are related to the
sample parameters and can be improved when new samples are
fabricated.

First, the smeared shape of the highest plateaus in
Fig.~\ref{IVcurves} is associated with the insufficient speed of
the device. This behavior of the sample is understood: For
$R=40$~k$\Omega$ yielding the value $(2z'-1) \approx 2$, the
tunneling rate at small $E$ was rather low, $\Gamma(E) \propto
E^2$. That is why a reliable ($p_{\rm CP}\rightarrow 1$) charge
transfer, governed by Eq.(\ref{Gamma}), was possible only at
significant values of $E \gtrsim E_{\rm{c}}$, achieved at
voltages $|V| \gtrsim V_0/4 \approx 170 ~\mu$V, and a rather low
frequency of the drive $f$. At somewhat lower resistance values,
viz. $R = 20-30$~k$\Omega$ yielding $(2z'-1)\approx 1$ and, hence,
$\Gamma_{\rm{j}}(E) \propto E$, correct operation should be
possible at smaller voltages, 170~$\mu$V~$\gtrsim |V|\gtrsim
E_{\rm{J}}/e \sim 16 ~\mu$V. As a result, wider and flatter
plateaus might appear. A room for increase in the operating
frequency is also available: The rate of the Landau-Zener process
leading to missing the pair transition, is evaluated at the
centers of the plateaus for our experimental values of
$E_{\rm{c}}$ and $E_{\rm{J}}$ as $p_{\rm LZ} = \exp(-f_0/f)$ with
$f_0 \approx 400$~MHz. This yields the values of $p_{\rm LZ}$
below $10^{-8}$ at $f \leq 20$~MHz. At smaller voltage bias
(resulting in smaller values of the derivative $\dot{E}$), this
type of operational errors might even be smaller. According to
Eq.(\ref{Relax}), the similar estimates are applicable to the
errors $p_{\rm err} = 1-p_{\rm rel}\approx p_{\rm LZ}$,
associated with the false transfer of pairs because of the
insufficient rate of relaxation.

Secondly, the rate of quasiparticle tunneling during operation
was still noticeable. One reason of this was the relatively large
value of the charging energy, $E_{\rm{c}} > \Delta_{\rm Al}$,
with the result that the even-odd parity blockade of quasiparticle
tunneling \cite{Tuominen} was not possible in this sample.
Furthermore, at the bias voltage $V \sim \pm E_{\rm{c}}/e$, the
voltages across individual junctions clearly exceeded the
threshold values at which the quasiparticle-current onset at the
double-gap voltage dramatically increased the tunneling rate. The
quasiparticle tunneling was, in our opinion, the main mechanism
leading to somewhat smaller values of the average current on the
plateaus. We interpret this behavior as a result of the sporadic
tunneling of charge $\pm e$, shifting the position of the
operation trajectory along the $V_{\rm g}$ axes and leading to
"the empty" cycles.

Although an Al-Cr sample with a smaller $E_{\rm{c}}$ value might
in this case be advantageous, a substantial improvement of the
shape of the plateaus appears to be possible in, e.g., Nb-Cr
samples. In such a sample $E_{\rm{c}}$ can be increased up to
$\sim\Delta_{\rm Nb}\approx 1.4$~meV (the energy gap of niobium),
keeping the quasiparticle tunneling totally suppressed. In this
case, the errors associated with the supercurrent (see
Eq.(\ref{GammaTot})) can also be reduced as low as $\sim 10^{-8}$
at $|V| \lesssim 0.06 V_0 \approx 80~\mu$V even if the electron
temperature in the resistor is as high as 100-200~mK.

To conclude, we predicted and experimentally demonstrated the
behavior of the dissipative system (the Al Bloch transistor with
attached Cr resistor) experiencing the adiabatic energy-level
crossing. Based on a strong dissimilarity in the probabilities to
pass the avoided crossing region along the lower and upper energy
branches, the remarkable unidirectional traversing of single
Cooper pairs across the transistor was realized when the gate
voltage was cycled. Similar circuits made from Nb (instead of Al)
may probably transfer single pairs with a relative accuracy of
$10^{-8}$ and, hence, meet the requirement of fundamental
metrology aiming at the construction of current and capacitance
standards \cite{MetrObz,MKeller}.

\section{acknowledgments}

This work is supported in part by the EU (Project COUNT).

\end{document}